\documentclass[a4paper]{article}

\usepackage{graphicx}
\usepackage{amsmath,amssymb}

\begin{document}
\setlength{\baselineskip}{18pt}

\title{\bf Entanglement entropies of coupled harmonic oscillators}

\author{Koichi Nakagawa \\[3mm] \small Laboratory of Physics,  \small School of Pharmacy and Pharmaceutical Sciences, \\ \small Hoshi University, Tokyo 142-8501, Japan}

\date{\empty}

\maketitle

\begin{abstract}

We study the quantum entanglement of systems of coupled harmonic oscillators on the basis of thermo-field dynamics (TFD). For coupled harmonic oscillators at equilibrium, the extended entanglement entropy is derived using the TFD method, and it is demonstrated to be controlled by temperature and coupling parameters.
For non-equilibrium systems, in addition to temperature and coupling parameters, the time dependence of the extended entanglement entropy is calculated in accordance with the dissipative von Neumann equation, and the dissipative dynamics of the systems of coupled harmonic oscillators is discussed. Consequently, based on TFD, the physical parametrization of the entanglement entropies is confirmed in both the equilibrium and non-equilibrium cases of harmonic oscillator systems by means of the laws of thermodynamics.

\end{abstract}
\vspace{5mm}

\section{Introduction}
\label{sec:1}

Entanglement has challenged our understanding of physics science the 1930s \cite{Schrodinger,EPR}, and it remains a highly relevant subject today. Since the development of quantum information science~\cite{Benett} in the last few decades, entanglement has acquired new significance. In quantum computation, entanglement is at the center of various computational processes, which cannot be performed classically. In particular, quantum communication protocols, such as quantum cryptography~\cite{Eckert}, quantum dense coding~\cite{Ben2}, quantum computation algorithms~\cite{Shor} and quantum state teleportation~\cite{Yurke2,Ben1,Bose}, can be understood using entangled states.
With regard to these papers, many studies on entanglement and its basic measure have been conducted.
It is well known that these are expressed by the measure of entanglement \cite{Ben3,Coffman} and concurrence \cite{Wootters,Rungta}.
Entangled quantum systems generally indicate correlations that are inherent in quantum theory, and entanglement in an aggregation of states is a signature of non-classicality \cite{Markham}.
In recent years, it has become clear that quantum information theory may connect with other fields of physics more deeply~\cite{Bargatin}.

On the other hand, the physical models of coupled harmonic oscillators have been used in many fields of physics, such as the Lee model in quantum field theory~\cite{Schweber,Fetter,Han,Dirac,Caves,Kim,Iachello}. There also exist models in which one of the variables is not observed~\cite{Fano, Prigogine, Takahashi,Feynman,Yurke,Eker,HKN}. These physical models are examples of Feynman's rest of the universe~\cite{Feynman,HKN99}. When a pair of coupled harmonic oscillators exists, one corresponds to the system in which we are interested, and the other corresponds to the rest of universe, which our measurement process does not reach. This implies that the system consists of two identical oscillators coupled together by an interaction term.

Motivated by Ref. \cite{Hashizume}, we began to develop a new attack to study quantum entanglement using thermo-field dynamics (TFD). In this new treatment of quantum entanglement with TFD, an extended density matrix has been defined in double Hilbert space (ordinary and conjugated Hilbert spaces), and some simple cases in two-spin systems have been examined~\cite{Hashizume,Nakagawa}. The new TFD-based method allows the entanglement states to be easily understood because the states in the ancillary Hilbert space play the role of a pursuer of the initial states. In the new formulation using TFD, it is important that the extended density matrix, $\hat{\rho}$, can be defined in the dual Hilbert space. We can therefore introduce the extended density matrix as $\hat{\rho}_{\rm A}:=\text{Tr}_{\rm B} \hat{\rho}$. We then obtain the extended entanglement entropy, $\hat{S}_{\rm A}$, for an equilibrium system as $\hat{S}_{\rm A}:=-k_B\text{Tr}_{\rm A}[\hat{\rho}_{\rm A}\log \hat{\rho}_{\rm A}]$. Moreover, by the general representation theorem (GRT) of TFD [36, 37], it has been shown that the TFD method is applicable not only to equilibrium systems but also to non-equilibrium systems [31]. In Ref. [31], the extended density matrix, $\hat{\rho}(t)$, in non-equilibrium systems was introduced with the help of the von Neumann equation [34, 35]. In non-equilibrium systems, therefore, the extended entanglement entropy, $\hat{S}_{\rm A}(t)$, can be defined as $\hat{S}_{\rm A}(t):=-k_B\text{Tr}_{\rm A}[\hat{\rho}_{\rm A}(t)\log \hat{\rho}_{\rm A}(t)]$. In Refs. [32, 33], indeed, $\hat{S}_{\rm A}(t)$ has been calculated and analyzed for two-spin systems. Additionally, it is also important to mention that, in two-spin systems at finite temperatures, the TFD method is useful for separating the fluctuations caused by quantum entanglement, i.e., the thermal and classical fluctuations can be separated.

It may be appropriate to apply the TFD method to systems of coupled harmonic oscillators. As discussed in the subsequent section, the Hamiltonian of the coupled harmonic oscillators can be expressed by the diagonalized, infinite matrix. Thus, the states of the coupled harmonic oscillators do not seem to do time evolution, however, do when dissipation enter. Indeed, it can be seen that the time evolution by the dissipation expresses the transitive process from a equilibrium state at $t=0$ to a equilibrium state at $t\to \infty$.The transitive process can be regarded as a type of non-equilibrium process in coupled harmonic oscillators.
In the present communication, therefore, we consider the above non- equilibrium process by reason of GRT of TFD [36, 37] and dissipative von Neumann equation [34, 35].
For the equilibrium systems of coupled harmonic oscillators, we derive the extended entanglement entropy using the TFD method and demonstrate that it is controlled by temperature and coupling parameters. 
For non-equilibrium systems, in addition to temperature and coupling parameters, we calculate the time dependence of the extended entanglement entropy in accordance with the dissipative von Neumann equation [34, 35], and we discuss the dissipative dynamics of the systems of coupled harmonic oscillators. Consequently, based on TFD, we can confirm the physical parametrization of the entanglement entropies in both equilibrium and non-equilibrium cases of harmonic oscillator systems by means of the laws of thermodynamics.

The rest of this paper is organized as follows. In the next section, we introduce the model of coupled harmonic oscillators and examine equilibrium systems at the ground state. In Sec.\,3, we calculate the extended entanglement entropy of the equilibrium systems of coupled harmonic oscillators with a heat bath, and we discuss the numerical results of temperature and coupling-parameter dependences. In Sec.\,4, by solving the dissipative von Neumann equation, we obtain the extended entanglement entropy of the non-equilibrium systems of coupled harmonic oscillators with a heat bath, and we discuss the numerical results of time, temperature, and coupling-parameter dependences. From the results, the dissipative non-equilibrium dynamics of the oscillator systems is argued. Finally, conclusions
are presented in Sec.\,5.

\section{Quantum mechanics of coupled harmonic oscillators} 
Let us consider a system of two coupled harmonic oscillators parametrized by the planar coordinates $(X_{\rm A}, X_{\rm B})$, momenta $(P_{\rm A}, P_{\rm B})$, and masses $(m_{\rm A},m_{\rm B})$. According to Ref. \cite{HKN99}, its Hamiltonian can be written as
\begin{align}\label{HAM1}
H &:= \frac{{P_{\rm A}}^2}{2m_{\rm A}} + \frac{{P_{\rm B}}^2}{2m_{\rm B}} + \frac{1}{2}\left( C_1 {X_{\rm A}}^2 + C_2 {X_{\rm B}}^2 + C_3 X_{\rm A} X_{\rm B} \right),
\end{align}
where $C_1, C_2$, and $C_3$ are constant parameters.
By using the rescaling transformations of the coordinates $\left( x_{\rm A},~x_{\rm B} \right) =\left( \mu X_{\rm A},~\mu ^{-1}X_{\rm B} \right) $, and the momenta $\left( p_{\rm A},~p_{\rm B} \right) =\left( \mu ^{-1}P_{\rm A},~\mu P_{\rm B} \right) $, the Hamiltonian \eqref{HAM1} can be expressed as
\begin{align}\label{HAM2}
H &:= H_F +H_I,\\
\label{HAM2'}
H_F &:=\frac{1}{2m}\left({p_{\rm A}}^2 + {p_{\rm B}}^2 \right),\qquad \mbox{and}\qquad H_I := \frac{1}{2}\left( c_1 {x_{\rm A}}^2 + c_2 {x_{\rm B}}^2 + c_3 x_{\rm A} x_{\rm B} \right),
\end{align}
where $\mu=({m_{1}/ m_{2}})^{1/ 4},~m = (m_{1}m_{2})^{1/2},~c_1=C_1\sqrt{m_2\over m_1},~c_2=C_2\sqrt{m_1\over m_2},~\mbox{and}~c_3=C_3$.
Because the Hamiltonian \eqref{HAM2} involves the interaction
term, $H_I$, in Eq. \eqref{HAM2'}, the straightforward analysis of the basic features of this system is not easy.
To simplify $H_I$, we express it in a matrix form as follows:
\begin{align}
H_I = \sum _{j=\rm A,B}\sum _{\ell =\rm A,B}x_{j}M_{j\ell }x_{\ell },
\end{align}
where
\begin{align}
\left( M_{j\ell } \right) = \left(
\begin{array}{cc}
 c_1 & \frac{c_3}{2} \\
 \frac{c_3}{2} & c_2 \\
\end{array}
\right).
\end{align}
Then, in terms of a transformation matrix,
\begin{equation} \label{UMAT}
(D_{j\ell }) =
\begin{pmatrix}
 \frac{c_1-c_2-\sqrt{\left(c_1-c_2\right){}^2+{c_3}^2}}{\sqrt{2} c_3} & \frac{1}{\sqrt{2}} \\[3mm]
 \frac{c_1-c_2+\sqrt{\left(c_1-c_2\right){}^2+{c_3}^2}}{\sqrt{2} c_3} & \frac{1}{\sqrt{2}} \\
\end{pmatrix},
\end{equation}
we can diagonalize $\left( M_{j\ell } \right)$ as
\begin{equation}
(D_{j\ell }^{-1}) (M_{\ell m})(D_{mn })=\left(
\begin{array}{cc}
 \frac{c_1+c_2-\sqrt{\left(c_1-c_2\right){}^2+{c_3}^2}}{2} & 0 \\
 0 & \frac{c_1+c_2+\sqrt{\left(c_1-c_2\right){}^2+{c_3}^2}}{2} \\
\end{array}
\right)
.
\end{equation}
Here, in order that the matrix, $(D_{j\ell })$, satisfy the orthogonality relation, $(^tD_{j\ell })(D_{\ell m})=\left( \delta_{jm} \right) $, we must set the parameters as $c_1=c_2$. It is noteworthy that the three constant parameters can be reduced to two by the orthogonality condition $c_1=c_2$.
This parametrization corresponds to that of the mixing angle in Ref.\,\cite{Jellal}, which is taken as $\pi /2$.
This implies that the system consists of two identical oscillators coupled together by an interaction term.
Through the transformation to the new space coordinates,
\begin{equation}
\begin{pmatrix}
y_{\rm A} \\[3mm]
y_{\rm B} \\
\end{pmatrix}
=
\begin{pmatrix}
 \frac{-{\rm sgn}{\left( c_3 \right) }}{\sqrt{2}} & \frac{1}{\sqrt{2}} \\[3mm]
 \frac{{\rm sgn}{\left( c_3 \right) }}{\sqrt{2}} & \frac{1}{\sqrt{2}} \\
\end{pmatrix}
\begin{pmatrix}
x_{\rm A} \\[3mm]
x_{\rm B} \\
\end{pmatrix},
\end{equation}
therefore, we obtain the decoupled Hamiltonian:
\begin{align}
\label{HAM3} H &= \frac{1}{2m} \left({p_{\rm A}}^2 +
{p_{\rm B}}^2 \right) +
\frac{k}{2}\left(e^{-2\eta } {y_{\rm A}}^2 + e^{2\eta }
{y_{\rm B}}^2\right),
\end{align}
where two parameters,
\begin{equation}\label{PARA}
k = \sqrt{{c_1}^2 - \frac{{c_3}^{2}}{4}}  \qquad \mbox{and}\qquad  e^{2\eta}= \frac {2c_1 +|c_3|}{2k},
\end{equation}
have been introduced. As can be seen from the decoupled Hamiltonian in Eq. \eqref {HAM3}, the two parameters, $k$ and $e^{2\eta}$, correspond to the effective spring constant and effective coupling constant, respectively.

Furthermore, the Hamiltonian \eqref{HAM3} can simply be diagonalized by defining a set of annihilation and creation operators,
\begin{equation}
\label{CRAN}
a_j = \sqrt{\frac{k}{2\hslash\omega}} e^{-\frac{\varepsilon\eta}{2}}y_j +\frac{i}{\sqrt{2m\hslash\omega}} e^{\frac{\varepsilon\eta}{2}}\hat{p}_j, \qquad\mbox{and}
\qquad a_j^{\dagger} = \sqrt{\frac{k}{2\hslash\omega}} e^{-\frac{\varepsilon\eta}{2}}y_j
- {\frac{i}{\sqrt{2m\hslash\omega}}}
e^{\frac{\varepsilon\eta}{2}}\hat{p}_j,
\end{equation}
respectively, with {the} frequency parameter,
\begin{equation}
\omega=\sqrt{\frac{k}{m}},
\end{equation}
{and} $\varepsilon=\pm 1$ for $j=\rm A,B$, respectively.
They satisfy the commutation relations
\begin{equation}
 [a_j, a_{\ell }^{\dagger}] = \delta_{j\ell },
\end{equation}
whereas other commutators vanish. Now we can map $H$ in terms of
$a_j$ and $a_j^{\dagger}$ as
\begin{equation}
\label{HAM7} H= \hslash\omega \left( e^{-\eta }a_{\rm A}^{\dagger}a_{\rm A} + e^{\eta}a_{\rm B}^{\dagger}a_{\rm B} +
\cosh\eta\right).
\end{equation}

To obtain the eigenstates and eigenvalues, one
solves the eigenvalue equation,
\begin{equation}
 H|n_{\rm A}, n_{\rm B}\rangle = {\cal E}_{n_{\rm A},n_{\rm B}} |n_{\rm A},n_{\rm B}\rangle,
\end{equation}
to obtain the states
\begin{equation}\label{frst}
|n_{\rm A},n_{\rm B}\rangle= \frac{(a_{\rm A}^{\dagger})^{n_{\rm A}} (a_{\rm B}^{\dagger})^{n_{\rm B}} }{\sqrt{n_{\rm A}!n_{\rm B}!}} |0, 0\rangle,
\end{equation}
as well as the energy spectrum,
\begin{equation}
\label{SPE2} {\cal E}_{n_{\rm A},n_{\rm B}} = {\hslash\omega} \left(e^{-\eta}n_{\rm A} +e^{\eta} n_{\rm B} +\cosh\eta\right).
\end{equation}
It is clear that these eigenvalues reduce to {those of} ordinary harmonic oscillators, ${\hslash\omega} \left(n_{\rm A} +  n_{\rm B} +1\right)$, at $\eta \to 0$. It is also worth noting that, although $H$ in Eq.\,\eqref{HAM7} looks to be decoupled in $a_{\rm A}$ and $a_{\rm B}$ oscillator modes, this is not in the original oscillator modes, i.e., the ground state of the coupled harmonic oscillators is actually entangled. Moreover, Eq.\,\eqref{SPE2} clearly shows that the existence of the parameters, $\omega $ and $\eta$, make a contribution and allow us to derive interesting results in the forthcoming analysis.

\section{Equilibrium systems with heat bath}


Let us consider the excited states $|n_{\rm A},~n_{\rm B}\rangle$ to the ground state in Eq.\,\eqref{frst}.
Of course, the states satisfy the ortho-normal relation $\langle m_{\rm A},~m_{\rm B}|n_{\rm A},~n_{\rm B}\rangle =\delta _{m_{\rm A},n_{\rm A}}\delta _{m_{\rm B},n_{\rm B}}$.
By using the base $\left\{ |n_{\rm A},~n_{\rm B}\rangle\right\} $, the matrix form of the diagonalized Hamiltonian \eqref{HAM7} is expressed as
\begin{equation}\label{eq:20}
H=\hslash\omega \sum _{n_{\rm A}=0}^{\infty }\sum _{n_{\rm B}=0}^{\infty }\left(e^{-\eta}n_{\rm A} +e^{\eta} n_{\rm B} +\cosh\eta\right)|n_{\rm A},~n_{\rm B}\rangle \langle n_{\rm A},~n_{\rm B}|.
\end{equation}
By using the diagonal Hamiltonian \eqref{eq:20}, the partition function of the systems is given by
\begin{align}
&Z(K ,\eta ):={\rm Tr_{A,B}}e^{-\beta H}=\frac{\exp \left( -K\cosh \eta \right) }{\left( 1-\exp \left( -K e^{-\eta} \right) \right) \left( 1-\exp \left( -K e^{\eta} \right) \right)}
,
\end{align}
where $\beta $ is the inverse temperature and $K=\hslash\omega\beta$ is the scaled inverse temperature.
In this case, the ordinary density matrix, $\rho _{\rm eq}(K ,\eta ):={e}^{-\beta H}/Z(K ,\eta )$, is obtained as follows:
\begin{align}
\rho _{\rm eq}(K ,\eta )=\frac{1}{Z(K ,\eta )}\sum _{n_{\rm A}=0}^{\infty }\sum _{n_{\rm B}=0}^{\infty }\exp \left( -K\left(e^{-\eta}n_{\rm A} +e^{\eta} n_{\rm B} +\cosh\eta\right) \right) |n_{\rm A},~n_{\rm B}\rangle \langle n_{\rm A},~n_{\rm B}|.\label{eq:22}
\end{align}
It is notable that the frequency parameter, $\omega $, has been absorbed into the parameter $K$. We proceed by claiming that the parameters, $K$ and $\eta $, will be used to study the entanglement in the present systems.

In the TFD formulation of the double Hilbert space, the statistical state, $|\Psi \rangle $, was originally defined as
$
|\Psi \rangle :=\sum _n \rho ^{1/2}|n \rangle |\tilde{n} \rangle ,
$
with the eigenstates, $\left\{ |n \rangle  \right\} $, and their copies, $\left\{ |\tilde{n} \rangle  \right\} $, where $\rho $ is the ordinary density matrix. Moreover, the extended density matrix, $\hat{\rho }$, was defined as
$
\hat{\rho }:=|\Psi \rangle \langle \Psi |
$.
In the present case, by using Eq. \eqref{eq:22}, the statistical state, $|\Psi (K ,\eta )\rangle $, and the extended density matrix, $\hat{\rho }(K ,\eta )$, are then reduced to
\begin{align}
|\Psi (K ,\eta )\rangle &=\sum _{n_{\rm A}=0}^{\infty }\sum _{n_{\rm B}=0}^{\infty }\rho _{\rm eq}(K ,\eta )^{1/2} |n_{\rm A},~n_{\rm B}\rangle |\tilde{n}_{\rm A},~\tilde{n}_{\rm B}\rangle \nonumber \\
&=\frac{1}{\sqrt{Z(K ,\eta )}}\sum _{n_{\rm A}=0}^{\infty }\sum _{n_{\rm B}=0}^{\infty }\exp \left( \frac{-K\left(e^{-\eta}n_{\rm A} +e^{\eta} n_{\rm B} +\cosh\eta\right)}{ 2 } \right) |n_{\rm A},~n_{\rm B}\rangle |\tilde{n}_{\rm A},~\tilde{n}_{\rm B}\rangle ,
\end{align}
and
\begin{align}
\hat{\rho }(K ,\eta )&=|\Psi (K ,\eta )\rangle \langle \Psi (K ,\eta )|\nonumber \\
&=\frac{1}{Z(K ,\eta )}\sum _{m_{\rm A}=0}^{\infty }\sum _{m_{\rm B}=0}^{\infty }\sum _{n_{\rm A}=0}^{\infty }\sum _{n_{\rm B}=0}^{\infty }\exp \left( \frac{-K\left(e^{-\eta}\left( m_{\rm A}+n_{\rm A} \right)  +e^{\eta} \left( m_{\rm B}+n_{\rm B} \right) +2\cosh\eta\right)}{ 2 } \right) \nonumber \\
&\hspace{13mm}\times |m_{\rm A},~m_{\rm B}\rangle \langle n_{\rm A},~n_{\rm B}||\tilde{m}_{\rm A},~\tilde{m}_{\rm B}\rangle  \langle \tilde{n}_{\rm A},~\tilde{n}_{\rm B}|,
\end{align}
respectively.
We then obtain the renormalized extended density matrix, $\hat{\rho }_{\rm A}(K ,\eta ):={\rm Tr_{B}}\hat{\rho }(K ,\eta )$, as follows:
\begin{align}
\hat{\rho }_{\rm A}(K ,\eta )&=\sum _{\ell _{\rm B}=0}^{\infty }\sum _{\tilde{\ell }_{\rm B}=0}^{\infty }\langle \ell _{\rm B}|\langle \tilde{\ell }_{\rm B}|\hat{\rho }(K ,\eta )|\ell _{\rm B}\rangle |\tilde{\ell }_{\rm B}\rangle \nonumber \\
&=\sum _{m_{\rm A}=0}^{\infty }\sum _{n_{\rm A}=0}^{\infty }\frac{\exp \left( -K\cosh \eta \right) \exp \left( \frac{-K e^{-\eta}\left( m_{\rm A}+n_{\rm A} \right) }{ 2 } \right) }{Z(K ,\eta )\left( 1-\exp \left( -K e^{\eta} \right) \right)} |m_{\rm A}\rangle \langle n_{\rm A}| |\tilde{m}_{\rm A}\rangle \langle \tilde{n}_{\rm A}|.
\end{align}

The extended entanglement entropy, $\hat{S}_{\rm A}:=-k_{\text{B}}\mathrm{Tr}_{\text{A}}\left[ \hat{\rho }_{\rm A}\log \hat{\rho }_{\rm A} \right]
$, in the present case is then reduced to
\begin{align}
\hat{S}_{\rm A}(K,\eta )&= \frac{k_{\text{B}} \left( K e^{-\eta} \left( \exp \left( \frac{K e^{-\eta} }{ 2 } \right) +\exp \left( K e^{-\eta} \right) \right)-\left( \exp \left( K e^{-\eta} \right)-1 \right) \log \left( \exp \left( K e^{-\eta} \right)-1 \right) \right) }{\left( \exp \left( \frac{K e^{-\eta} }{ 2 } \right)-1 \right)^2}.
\label{eq:29}
\end{align}
\begin{figure}
\begin{center}
\includegraphics[width=7cm]{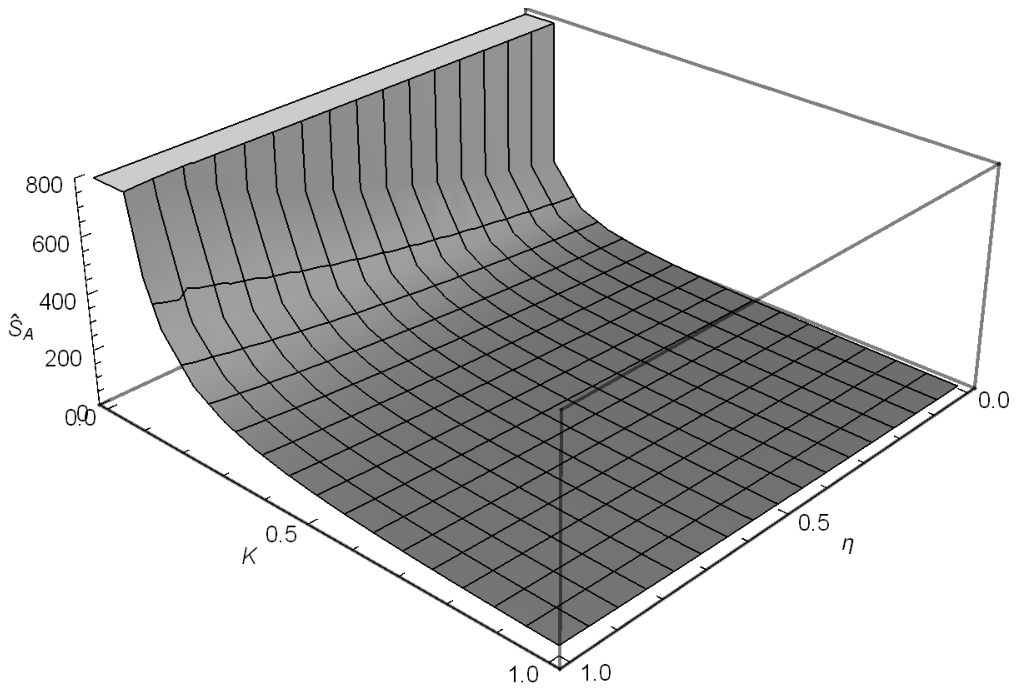} \includegraphics[width=7cm]{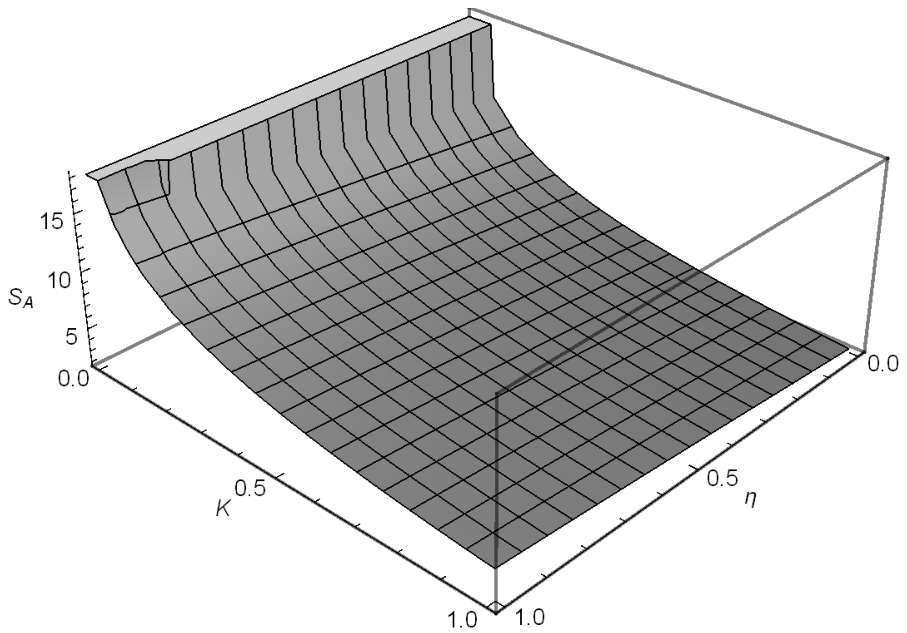}
\caption{The $K-\eta $ distribution of the extended entanglement entropy,  $\hat{S}_{\rm A}(K,\eta )$, in Eq. \eqref{eq:29} and the ordinary entanglement entropy, $S_{\rm A}(K,\eta )$.}\label{fig:2}
\end{center}
\end{figure}
The $K-\eta $ distribution of $\hat{S}_{\rm A}(K,\eta )$ is shown in Fig.\,\ref{fig:2} (in units of $k_{\text{B}}=1$), along with the ordinary entanglement entropy, $S_{\rm A}(K,\eta )$, of the coupled harmonic oscillators. As can clearly be visualized from Fig.\,\ref{fig:2}, $\hat{S}_{\rm A}(K,\eta )$ decreases with respect to $K$ and increases with respect to $\eta $. In addition, $\hat{S}_{\rm A}(K,\eta )$ is positive and vanishes at zero temperature. It is worth noting that $S_{\rm A}(K,\eta )$ satisfies Nernst's theorem. Thus, this result can be interpreted as indicating that the parametrization with $K$ and $\eta $ is the physically reasonable one in the equilibrium systems.  It is also interesting to note that, in the region of $(K,\eta )$, $\hat{S}_{\rm A}(K,\eta )>S_{\rm A}(K,\eta )$. This result can be interpreted to imply that the degrees of freedom of the dual Hilbert space  are larger than those of the physical Hilbert space, according to the letter.

\section{Non-equilibrium systems with heat bath}

\begin{figure}[hbpt]
\begin{center}
\unitlength 1mm
\begin{picture}(100,80)
\put(0,0){\includegraphics[width=10cm]{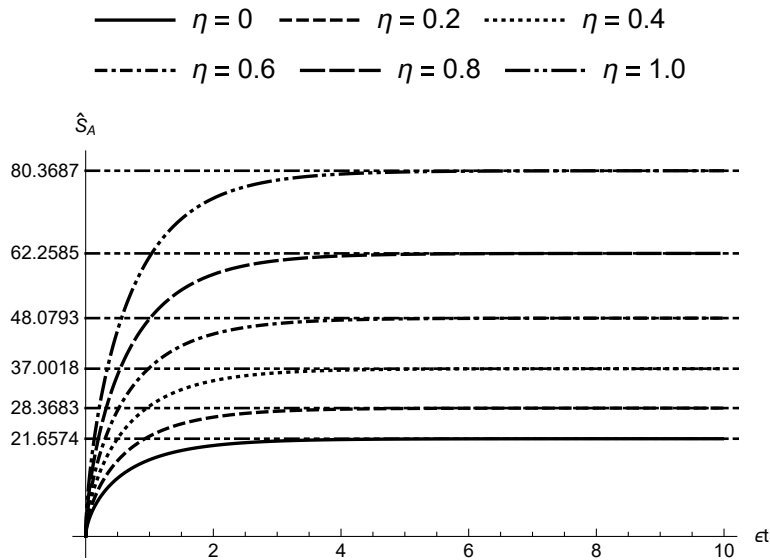}}
\end{picture}
\end{center}
\caption{Temperature dependence of $\hat{S}_{\rm A}(K,\eta ;t)$ in Eq. \eqref{eq:48} in the non-equilibrium system with $K=0.5$ and $\epsilon t \sim 10$. The values on the vertical axis represent the asymptotic values given by Eq.\,\eqref{eq:37a} (left) and the ordinary entanglement entropy, $S_{\rm A}(K,\eta ;t)$, (right).}\label{fig:03}
\end{figure}

Now, we consider the Hamiltonian in Eq.\,\eqref{eq:20} as a non-equilibrium system with dissipations. To include the effects of dissipation, we adopt the dissipative von Neumann equation [34, 35]:
\begin{align}
i\hslash \frac{\partial }{\partial t}\rho (t)=\left[ H,\rho (t) \right] -\epsilon\left( \rho (t)-\rho _{\rm eq} \right) ,
\label{eq:30}
\end{align}
where $\epsilon$ is a dissipation parameter and $\rho _{\rm eq}$ was given by Eq.\,\eqref{eq:22}, as discussed in the previous section. The solution of Eq. \eqref{eq:30} is written as
\begin{align}
\rho (t)=e^{-\epsilon t}U^{\dagger }(t)\rho _0U(t)+\left( 1-e^{-\epsilon t} \right) \rho _{\rm eq} ,
\label{eq:31}
\end{align}
for an arbitrary density matrix, $\rho _0$, where the unitary operator, $U(t):=e^{iHt/\hslash}$, is expressed as
\begin{align}
U(t)=\sum _{n_{\rm A}=0}^{\infty }\sum _{n_{\rm B}=0}^{\infty }\exp \left( i\omega t \left(e^{-\eta}n_{\rm A} +e^{\eta} n_{\rm B} +\cosh\eta\right) \right) |n_{\rm A},~n_{\rm B}\rangle \langle n_{\rm A},~n_{\rm B}|.
\label{eq:32}
\end{align}
Because the explicit expression for $\rho (t)$ in Eq.\,\eqref{eq:31} is complicated for an arbitrary initial condition, we will hereafter confine the discussion to the initial condition
\begin{align}
\rho_0= |0,0\rangle\langle 0,0|.
\label{eq:33}
\end{align}
By inserting Eqs.\,\eqref{eq:22} and \eqref{eq:32}, along with the initial condition \eqref{eq:33}, into Eq.\,\eqref{eq:31}, we obtain the ordinary density matrix:
\begin{align}
\rho _{\rm neq}(K,\eta ;t)=\frac{1}{Z(K ,\eta )} \sum _{n_{\rm A}=0}^{\infty }\sum _{n_{\rm B}=0}^{\infty }& \left( \left( \frac{\delta _{n_{\rm A},0}\delta _{n_{\rm B},0}}{\left(1-\exp\left( -K e^{-\eta} \right) \right) \left(1-\exp\left( -K e^{\eta} \right) \right) }-1 \right) e^{-\epsilon t}+1 \right) \nonumber \\
&\times \exp \left( -K\left(e^{-\eta}n_{\rm A} +e^{\eta} n_{\rm B} +\cosh\eta\right) \right) |n_{\rm A},~n_{\rm B}\rangle \langle n_{\rm A},~n_{\rm B}|.
\label{eq:34}
\end{align}
The extended density matrix, $\hat{\rho }(t):=|\Psi (t)\rangle \langle \Psi (t)|$, is then reduced to
\begin{align}
\hat{\rho }(K,\eta ;t)&=\frac{1}{Z(K ,\eta )}\sum _{m_{\rm A}=0}^{\infty }\sum _{m_{\rm B}=0}^{\infty }\rho _{\rm neq}(K,\eta ;t)^{1/2} |m_{\rm A},~m_{\rm B}\rangle |\tilde{m}_{\rm A},~\tilde{m}_{\rm B}\rangle \nonumber \\
&\hspace{14mm}\times \sum _{n_{\rm A}=0}^{\infty }\sum _{n_{\rm B}=0}^{\infty } \langle n_{\rm A},~n_{\rm B}| \langle \tilde{n}_{\rm A},~\tilde{n}_{\rm B}| \rho _{\rm neq}(K,\eta ;t)^{1/2} \nonumber \\
&=\frac{\exp \left( -K\cosh \eta \right)}{Z(K ,\eta )} \sum _{m_{\rm A}=0}^{\infty }\sum _{n_{\rm A}=0}^{\infty }b_{m_{\rm A},n_{\rm A}}(t)\exp \left( \frac{-K e^{-\eta}\left( m_{\rm A}+n_{\rm A} \right) }{ 2 } \right) |m_{\rm A}\rangle \langle n_{\rm A}| |\tilde{m}_{\rm A}\rangle \langle \tilde{n}_{\rm A}|,
\end{align}
where
\begin{align}
b_{m_{\rm A},n_{\rm A}}(t):&=\left( 1-e^{-\epsilon t} \right) \frac{\exp \left( -K e^{\eta} \right) }{1-\exp \left( -K e^{\eta} \right) } \notag\\ 
&+  \sqrt{\left(  \frac{\delta _{m_{\rm A},0}}{\left(1-\exp\left( -K e^{-\eta} \right) \right) \left(1-\exp\left( -K e^{\eta} \right) \right) }-1 \right) e^{-\epsilon t}+1 }\notag\\
&\times\sqrt{ \left( \frac{\delta _{n_{\rm A},0}}{\left(1-\exp\left( -K e^{-\eta} \right) \right) \left(1-\exp\left( -K e^{\eta} \right) \right) }-1 \right) e^{-\epsilon t}+1 } .
\end{align}
The extended entanglement entropy, $\hat{S}_{\rm A}(K,\eta ;t):=-k_{\text{B}}\mathrm{Tr}_{\text{A}}\left[ \hat{\rho }_{\rm A}(K,\eta ;t)\log \hat{\rho }_{\rm A}(K,\eta ;t) \right]
$, is eventually described as
\begin{align}
\hat{S}_{\rm A}&(K,\eta ;t)= -k_{\text{B}}\frac{\exp \left( -K\cosh \eta \right)}{Z(K ,\eta )} \left( c_0(t)\log \left( \frac{c_0(t)\exp \left( -K\cosh \eta \right)}{Z(K ,\eta )} \right) -\frac{c(t)K e^{-\eta} \exp \left( \frac{K e^{-\eta} }{ 2 } \right)}{2 \left(\exp \left( \frac{K e^{-\eta} }{ 2 } \right)-1\right)^2} \right. \nonumber \\
&+ \frac{c(t)}{\exp \left( \frac{K e^{-\eta} }{ 2 } \right)-1}\log \left( \frac{c(t)\exp \left( -K\cosh \eta \right)}{Z(K ,\eta )} \right) +\frac{-K e^{-\eta} \exp \left( \frac{K e^{-\eta} }{ 2 } \right)}{2 \left(\exp \left( \frac{K e^{-\eta} }{ 2 } \right)-1\right)^2}\left(\frac{2d(t)}{\exp \left( \frac{K e^{-\eta} }{ 2 } \right)-1}+c(t) \right) \nonumber \\
&\left. +\frac{c(t)}{\exp \left( \frac{K e^{-\eta} }{ 2 } \right)-1}\log \left( \frac{c(t)\exp \left( -K\cosh \eta \right)}{Z(K ,\eta )} \right)  +\frac{d(t)}{\left( \exp \left( \frac{K e^{-\eta} }{ 2 } \right)-1 \right) ^2}\log \left( \frac{d(t)\exp \left( -K\cosh \eta \right)}{Z(K ,\eta )} \right) \right),
\label{eq:48}
\end{align}
where
\begin{align}
&c_0(t):= \left( 1-\mathrm{e}^{-\epsilon t} \right) \frac{\exp \left(  -K e^{\eta} \right) }{1-\exp \left( -K e^{\eta} \right) } +\left| \left(  \frac{1}{\left(1-\exp\left( -K e^{-\eta} \right) \right) \left(1-\exp\left( -K e^{\eta} \right) \right) }-1 \right) e^{-\epsilon t}+1\right|,\\
&c(t):=\left( 1-\mathrm{e}^{-\epsilon t} \right) \frac{\exp \left(  -K e^{\eta} \right) }{1-\exp \left( -K e^{\eta} \right) } \notag \\
&\hspace{3em}+\sqrt{\left( 1- \mathrm{e}^{-\epsilon t} \right) \left( \left(  \frac{1}{\left(1-\exp\left( -K e^{-\eta} \right) \right) \left(1-\exp\left( -K e^{\eta} \right) \right) }-1 \right) e^{-\epsilon t}+1\right) } ,
\end{align}
and
\begin{align}
d(t):&=\left( 1- \mathrm{e}^{-\epsilon t}  \right) + \left( 1-\mathrm{e}^{-\epsilon t} \right) \frac{\exp \left(  -K e^{\eta} \right) }{1-\exp \left( -K e^{\eta} \right) } =\frac{1-\mathrm{e}^{-\epsilon t}}{1-\exp \left( -K e^{\eta} \right) } .\label{eq:36}
\end{align}
As can clearly be seen from Eqs.\,\eqref{eq:48}, $\sim$,  \eqref{eq:36}, the behaviour of $\hat{S}_{\rm A}(K,\eta ;t)$ at $t\to\infty $ reduces to
\begin{align}
\lim _{t\to\infty }\hat{S}_{\rm A}(K,\eta ;t)&= \frac{k_{\text{B}}\left( K e^{-\eta} \left( \exp \left( \frac{K e^{-\eta} }{ 2 } \right) +\exp \left( K e^{-\eta} \right) \right)-\left( \exp \left( K e^{-\eta} \right)-1 \right) \log \left( \exp \left( K e^{-\eta} \right)-1 \right) \right) }{\left( \exp \left( \frac{K e^{-\eta} }{ 2 } \right)-1 \right)^2} ,
\label{eq:37a}
\end{align}
which is eventually consistent with the equilibrium expression in Eq.\,\eqref{eq:29}.
In addition, the behaviour of $\hat{S}_{\rm A}(K,\eta ;t)$ at $t\to 0$ reduces to
$
\lim _{t\to 0}\hat{S}_{\rm A}(K,\eta ;t)=0.
$
The time dependences of $\hat{S}_{\rm A}(K,\eta ;t)$ are shown in Fig.~\ref{fig:03} (in units of $k_{\text{B}}=1$). It is apparent from these figures that $\hat{S}_{\rm A}(K,\eta ;t)$ increases with respect to $t$ and saturates at $t\to \infty$. These results clearly show that the behaviour of $\hat{S}_{\rm A}(K,\eta ;t)$ is consistent with the second law of thermodynamics. Thus, these results can be interpreted as indicating that the parametrization with $K,~\eta $ and $t$ is a physically reasonable one even in non-equilibrium systems.

\section{Conclusions }
\label{sec:4}

In this communication, we examined the extended entanglement entropy of systems of coupled harmonic oscillators in both equilibrium and non-equilibrium cases on the basis of the TFD formulation.
For the equilibrium systems, the extended entanglement entropy has been derived using TFD, and their temperature and coupling-parameter dependences have been explicitly calculated. The results show that $\hat{S}_{\rm A}(K,\eta )$ decreases with increasing $K$ and increases with increasing $\eta $. Moreover, $\hat{S}_{\rm A}(K,\eta )$ is positive and vanishes at zero temperature. These findings seem to be consistent with Nernst's theorem. From the numerical results, it has been proven that the amplitude of the extended entanglement entropy becomes larger than that of the ordinary entanglement entropy, according to the extension of the Hilbert space. We can expect that this observation is true even in a non- equilibrium system. For the non-equilibrium systems, the extended entanglement entropy has been computed using TFD. The results show that $\hat{S}_{\rm A}(K,\eta ;t)$ increases with $t$ and saturates at $t\to\infty $ (i.e., thermal equilibrium state), which is in good agreement with the second law of thermodynamics.

We have succeeded in demonstrating that the parametrization of the entanglement entropies of the systems of coupled harmonic oscillators with $K,~\eta $, and $t$ is physically reasonable in both the equilibrium and non-equilibrium cases on the basis of the general TFD algorithm.
Indeed, the simple device of coupled harmonic oscillators can serve as an analog computer for many quantum field theories and their models. We can extend the new TFD-based method developed in this paper to study the extended entanglement in other interacting systems that illustrate the Feynmans rest. It is especially interesting to analyze the interacting fields in various space-time, using the present method.



\end{document}